\documentclass[prc,twocolumn,superscriptaddress,superscriptfootnote]{revtex4}
\usepackage{amssymb}

\usepackage{amsmath,amsfonts,amssymb,epsfig,color}

\begin{document}

\title{Transition probabilities in the
$U(3,3)$ limit of the symplectic IVBM}
\author{H. G. Ganev}
\affiliation{Bogoliubov Laboratory of Theoretical Physics, Joint
Institute for Nuclear Research, \\
141980 Dubna, Moscow Region, Russia}
\affiliation{Institute of
Nuclear Research and Nuclear Energy, Bulgarian Academy of Sciences,
\\ Sofia 1784, Bulgaria}

\setcounter{MaxMatrixCols}{10}

\begin{abstract}
The tensor properties of the algebra generators are determined in
respect to the reduction chain $Sp(12,R) \supset U(3,3) \supset
U_{p}(3) \otimes \overline{U_{n}(3)}\supset U^{\ast}(3) \supset
O(3)$, which defines one of the dynamical symmetry limits of the
Interacting Vector Boson Model (IVBM). The symplectic basis
according to the considered chain is thus constructed and the action
of the $Sp(12,R)$ generators as transition operators between the
basis states is illustrated. The matrix elements of the $U(3,3)$
ladder operators in the so obtained symmetry-adapted basis are
given. The $U(3,3)$ limit of the model is further tested on the more
complicated and complex problem of reproducing the $B(E2)$
transition probabilities between the collective states of the ground
band in $^{104}Ru$, $^{192}Os$, $^{192}Pt$, and $^{194}Pt$ isotopes,
considered by many authors to be axially asymmetric. Additionally,
the excitation energies of the ground and $\gamma$ bands in
$^{104}Ru$ are calculated. The theoretical predictions are compared
with the experimental data and some other collective models which
accommodate the $\gamma-$rigid or $\gamma-$soft structures. The
obtained results reveal the applicability of the model for the
description of the collective properties of nuclei, exhibiting
axially asymmetric features.
\end{abstract}
\maketitle
 PACS {21.60.Fw, 23.20.-g, 21.10.Re, 27.80.+w, 27.60.+j}
%\end{start}

\section{Introduction}

Symmetry is an important concept in physics. In finite many-body
systems, it appears as time reversal, parity, and rotational
invariance, but also in the form of dynamical symmetries
\cite{DG}-\cite{RW}. In the algebraic models, the use of the
dynamical symmetries defined by a certain reduction chain of the
group of dynamical symmetry yields exact solutions for the
eigenvalues and eigenfunctions of the model Hamiltonian, which is
constructed from the invariant operators of the subgroups in the
chain. Many properties of atomic nuclei have been investigated using
such models, in which one obtains bands of collective states which
span irreducible representations of the corresponding dynamical
groups.

Something more, it is very simple and straightforward to calculate
the matrix elements of transition operators between the eigenstates
as both - the basis states and the operators - can be defined as
tensor operators in respect to the considered dynamical symmetry.
Then the calculation of matrix elements is simplified by the use of
a respective generalization of the Wigner-Eckart theorem, which
requires the calculation of the isoscalar factors and reduced matrix
elements. By definition such matrix elements give the transition
probabilities between the collective states attributed to the basis
states of the Hamiltonian.

The comparison of the experimental data with the calculated
transition probabilities is one of the best tests of the validity of
the considered algebraic model. With the aim of such applications of
one of the dynamical symmetries of the symplectic Interacting Vector
Boson Model (IVBM), we develop in this paper a practical
mathematical approach for explicit evaluation of the matrix elements
of transitional operators in the model.

The IVBM and its recent applications for the description of diverse
collective phenomena in the low-lying energy spectra (see, e.g., the
review article \cite{pepan}) exploit the symplectic algebraic
structures and the Sp(12,$R$) is used as a dynamical symmetry group.
Symplectic algebras and their substructures have been applied
extensively in the theory of nuclear structure
\cite{GL}-\cite{Vanagas}. They are used generally to describe
systems with a changing number of particles or excitation quanta and
in this way provide for larger representation spaces and richer
subalgebraic structures that can accommodate the more complex
structural effects as realized in nuclei with nucleon numbers that
lie far from the magic numbers of closed shells. In particular, the
model approach was adapted to incorporate the newly observed higher
collective states, both in the first positive and negative parity
bands \cite{Sp12U6} by considering the basis states as "yrast"
states for the different values of the number of bosons $N$ that
built them.

In Ref.\cite{TSIVBM} a new dynamical symmetry limit of the IVBM was
introduced, which seems to be appropriate for the description of
deformed even-even nuclei, exhibiting triaxial features. Usually, in
the geometrical approach the triaxial nuclear properties are
interpreted in terms of either the $\gamma$-unstable rotor model of
Wilets and Jean \cite{WJ} or the rigid triaxial rotor model (RTRM)
of Davydov \emph{et al.} \cite{DF}. An alternative description can
be achieved by exploiting the properties of the $SU^{\ast}(3)$
algebra introduced in Ref.\cite{TSIVBM} (and appearing also in the
context of IBM-2 \cite{PSDiep}). The latter is appropriate for
nuclei in which the one type of particles is particle-like and the
other is hole-like. Using a schematic Hamiltonian with a perturbed
$SU^{\ast}(3)$ dynamical symmetry, the IVBM was applied for the
calculation of the low-lying energy spectrum of the nucleus
$^{192}$Os \cite{TSIVBM}. The obtained results proved the relevance
of the proposed dynamical symmetry in the description of deformed
triaxial nuclei.

In this paper we develop further our theoretical approach initiated
in Ref.\cite{TSIVBM} by considering the transition probabilities in
the framework of the symplectic IVBM with $Sp(12,R)$ as a group of
dynamical symmetry. For this purpose we consider the tensorial
properties of the algebra generators in respect to the reduction
chain:
\begin{align}
\begin{tabular}{lllllll}
$Sp(12,R)$ & $\supset $ & $\left\{
\begin{tabular}{l}
$\ U(6)$ \\
$U(3,3)$%
\end{tabular}%
\right\} $ &  &  &  &  \\
&  &  &  &  &  &  \\
& $\supset $ & $U_{p}(3)\otimes \overline{U_{n}(3)}$ & $\supset $ & $U^{\ast }(3)$ & $%
\supset $ & $SO(3)$,%
\end{tabular}
\label{NL}
\end{align}
where $U_{p}(3)$ and $\overline{U_{n}(3)}$ are the one-fluid
algebras corresponding to the two nuclear subsystems, $U^{\ast}(3)$
is the combined two-fluid algebra, and $SO(3)$ is the standard
angular momentum algebra. Further we classify the basis states by
the quantum numbers corresponding to the irreducible representations
(irreps) of different subgroups along the chain (\ref{NL}). In this
way we are able to define the transition operators between the basis
states and then to evaluate analytically their matrix elements. This
will allow us further to test the model in the description of the
electromagnetic properties observed in some non-axial nuclei. As a
first step we will test the theory on the transitions between the
states belonging to the ground state bands (GSB) in some even-even
nuclei from the $A \approx 100$ and $A \approx 190$ mass regions.

\section{Tensorial properties of the generators of the Sp(12,R) group}

It was suggested by Bargmann and Moshinsky \cite{BargMosh} that two
types of bosons are needed for the description of nuclear dynamics.
It was shown there that the consideration of only two-body system
consisting of two different interacting vector particles will
suffice to give a complete description of $N$ three-dimensional
oscillators with a quadrupole-quadrupole interaction. The latter can
be considered as the underlying basis in the algebraic construction
of the \emph{phenomenological} IVBM.

The basic building blocks of the IVBM \cite{TSIVBM} are the creation
and annihilation operators of two kinds of vector bosons
$u_{m}^{\dag}(\alpha )$ and $u_{m}(\alpha )$ $(m=0,\pm 1)$, which
differ in an additional quantum number $\alpha=\pm1/2$ (or
$\alpha=p$ and $n$)$-$the projection of the $T$-spin (an analogue to
the $F$-spin of IBM-2 or the $I$-spin of the particle-hole IBM). In
the present paper, we consider these two bosons just as elementary
building blocks or quanta of elementary excitations (phonons) rather
than real fermion pairs, which generate a given type of algebraic
structures. Thus, only their tensorial structure is of importance
and they are used as an auxiliary tool, generating an appropriate
\emph{dynamical} symmetry.

The vector bosons can be considered as components of a
$6-$dimensional vector, which transform according to the fundamental
$U(6)$ irreducible representation $[1,0,0,0,0,0]_{6}\equiv [1]_{6}$
and its conjugate (contragradient) one $[0,0,0,0,0,-1]_{6}\equiv
[1]_{6}^{\ast }$, respectively. These irreducible representations
become reducible along the chain of subgroups (\ref{NL}) defining
the dynamical symmetry. This means that along with the quantum
number characterizing the representations of $U(6)$, the operators
are also characterized by the quantum numbers of the subgroups of
chain (\ref{NL}). Introducing the notations
$u_{i}^{\dag}(\frac{1}{2})=p_{i}^{\dag}$ and
$u_{i}^{\dag}(-\frac{1}{2})=n_{i}^{\dag}$, the components of the
creation operators $u_{m}^{\dag}(\alpha )$ labeled by the chain
(\ref{NL}) can be written as:
\begin{equation}
p^{\dag}_{m}\equiv p^{{\dag}[1]_{6}}_{[1]_{3}[0]^{\ast}_{3} \
[1]_{3}(1)_{3}m}, \ \ \ n^{\dag}_{m}\equiv
n^{{\dag}[1]_{6}}_{[0]_{3}[1]^{\ast}_{3} \ [1]^{\ast}_{3}(1)_{3}m}.
\label{basten}
\end{equation}

According to the chain (\ref{NL}), the fundamental $U(6)$ irrep
$[1]_{6}$ decomposes as
\begin{equation}
[1]_{6} \supset [1]_{3} \oplus [1]^{\ast}_{3}, \label{1NDec}
\end{equation}
i.e. as a direct product sum of the $U_{p}(3)$ and
$\overline{U_{n}(3)}$ fundamental irreps. In Eq.(\ref{1NDec}) the
$[1]^{\ast}_{3}$ denotes the (contragradient) irrep of
$\overline{U_{n}(3)}$ which is conjugate to the $[1]_{3}$ of
$U_{p}(3)$. This corresponds to the case when the one type of
particles in the two-fluid nuclear system is particle-like and the
other is hole-like. Note that there is an alternative decomposition
of the fundamental $U(6)$ irrep $[1]_{6}$:
\begin{equation}
[1]_{6} \supset [1]_{3} \oplus [1]_{3}, \label{1NDecPP}
\end{equation}
where the group $\overline{U_{n}(3)}$ in Eq.(\ref{NL}) should be
replaced by the $U_{n}(3)$ one. The decomposition (\ref{1NDecPP}) is
appropriate for the situation when the nucleus is considered as
consisting of two particle-like constituents. In our further
considerations we will need also the reduction of the $U(6)$ irrep
$[2]_{6}$ along the chain (\ref{NL}). According to the decomposition
rules for the fully symmetric $U(6)$ irreps, we obtain for the
$U_{p}(3)\otimes \overline{U_{n}(3)}$ content
\begin{equation}
[2]_{6} \supset [2]_{3}[0]^{\ast}_{3} + [1]_{3}[1]^{\ast}_{3} +
[0]_{3}[2]^{\ast}_{3}. \label{2NDec}
\end{equation}

Thus, the generators of the symplectic group $Sp(12,R)$ can already
be defined as irreducible tensor operators according to the whole
chain (\ref{NL})\ of subgroups as follows.

The raising operators of $Sp(12,R)$ can be expressed as
\begin{align}
F_{[\lambda]_{3}[0]_{3}^{\ast} \quad[\lambda]_{3}}^{\quad [\chi
]_{6}\quad LM}
&=C_{[1]_{3}[0]_{3}^{\ast}\quad[1]_{3}[0]^{\ast}_{3}\quad
[\lambda]_{3}[0]^{\ast}_{3}}^{[1]_{6}\quad \quad [1]_{6}\quad \quad
\quad [\chi]_{6}}C^{[\lambda]_{3}}_{[1]_{3},[1]_{3}} \notag \\
&\times C_{(1)_{3}(1)_{3}(L)_{3}}^{[1]_{3}[1]_{3}[\lambda]_{3}}
C_{1m1n}^{LM} \notag \\
&\times p^{{\dag}[1]_{6}}_{[1]_{3}[0]^{\ast}_{3} \ [1]_{3}(1)_{3}m}
p^{{\dag}[1]_{6}}_{[1]_{3}[0]^{\ast}_{3} \ [1]_{3}(1)_{3}n},
\label{Fpp}
\end{align}
\begin{align}
F_{[0]_{3}[\lambda]_{3}^{\ast} \quad[\lambda]^{\ast}_{3}}^{\quad
[\chi ]_{6}\quad LM}
&=C_{[0]_{3}[1]_{3}^{\ast}\quad[0]_{3}[1]^{\ast}_{3}\quad
[0]_{3}[\lambda]^{\ast}_{3}}^{[1]_{6}\quad \quad [1]_{6}\quad \quad
\quad [\chi]_{6}}C^{[\lambda]_{3}}_{[-1]_{3},[-1]_{3}} \notag \\
&\times
C_{(1)_{3}(1)_{3}(L)_{3}}^{[1]^{\ast}_{3}[1]^{\ast}_{3}[\lambda]^{\ast}_{3}}
C_{1m1n}^{LM} \notag \\
&\times n^{{\dag}[1]_{6}}_{[0]_{3}[1]^{\ast}_{3} \
[1]^{\ast}_{3}(1)_{3}m} n^{{\dag}[1]_{6}}_{[0]_{3}[1]^{\ast}_{3} \
[1]^{\ast}_{3}(1)_{3}n}, \label{Fnn}
\end{align}
\begin{align}
F_{[1]_{3}[1]_{3}^{\ast} \quad[\lambda]_{3}}^{\quad [\chi ]_{6}\quad
LM} &=C_{[1]_{3}[0]_{3}^{\ast}\quad[0]_{3}[1]^{\ast}_{3}\quad
[0]_{3}[\lambda]_{3}}^{[1]_{6}\quad \quad [1]_{6}\quad \quad
\quad [\chi]_{6}}C^{[\lambda]_{3}}_{[1]_{3},[-1]_{3}} \notag \\
&\times
C_{(1)_{3}(1)_{3}(L)_{3}}^{[1]_{3}[1]^{\ast}_{3}[\lambda]_{3}}
C_{1m1n}^{LM} \notag \\
&\times p^{{\dag}[1]_{6}}_{[1]_{3}[0]^{\ast}_{3} \ [1]_{3}(1)_{3}m}
n^{{\dag}[1]_{6}}_{[0]_{3}[1]^{\ast}_{3} \ [1]^{\ast}_{3}(1)_{3}n},
\label{Fpn}
\end{align}
where, according to the lemma of Racah \cite{Racah}, the
Clebsch-Gordan coefficients along the chain are factorized by means
of the isoscalar factors (IF), defined for each step of
decomposition (\ref{NL}). The lowering operators $G_{[\lambda'
]_{3}[\lambda'']_{3}\quad [\lambda]_{3}}^{\quad [\chi
]_{6}\quad\quad LM}$ of $Sp(12,R)$ are obtained from the rasing ones
$F_{[\lambda' ]_{3}[\lambda'']^{\dag}_{3}\quad [\lambda]_{3}}^{\quad
[\chi ]_{6}\quad\quad LM}$ by Hermition conjugation. That is why we
consider only the tensor properties of the raising operators.

The tensors (\ref{Fpp})-(\ref{Fpn}) transform according to
\begin{equation}
[1]_{6}\times \lbrack 1]_{6}=[2]_{6}+[1,1]_{6}, \label{ftd}
\end{equation}
and their Hermition conjugate counterparts according to
\begin{equation}
[1]_{6}^{\ast }\times \lbrack 1]_{6}^{\ast}=[-2]_{6}+[-1,-1]_{6},
\end{equation}
respectively. But, since the basis states of the IVBM are fully
symmetric, we consider only the fully symmetric $U(6)$
representation $[2]_{6}$ and its conjugate $[-2]_{6}$. Hence, the
tensors (\ref{Fpp})-(\ref{Fpn}) transform according to the $U(6)$
irrep $[\chi]_{6}\equiv [2]_{6}$.

The tensor (\ref{Fpp}) with respect to the $U^{\ast}(3)$ subgroup
transforms according to the direct product
\begin{equation}
[1]_{3} \times [1]_{3}=[2]_{3}+[1,1]_{3},
\end{equation}
while (\ref{Fnn}) and (\ref{Fpn}) transform according to
\begin{equation}
[1]^{\ast}_{3} \times [1]^{\ast}_{3}
=[2,2]_{3}+[2,1,1]_{3}=[-2]_{3}+[1]_{3},
\end{equation}
\begin{equation}
[1]_{3} \times [1]^{\ast}_{3}=[2,1]_{3} + [1,1,1]_{3}=
[1,-1]_{3}+[0]_{3}
\end{equation}
and obviously, because of their symmetric character, (\ref{Fpp}) and
(\ref{Fnn}) transform only according to the symmetric $U^{\ast}(3)$
representations $[2]_{3}$ and $[-2]_{3}$, respectively. The latter
follows also from the reduction (\ref{2NDec}). In this way we obtain
the following set of raising generators:
\begin{align}
&F_{[2]_{3}[0]_{3}^{\ast} \quad[2]_{3}}^{\quad [2]_{6}\quad LM},
\quad F_{[0]_{3}[2]_{3}^{\ast} \quad[-2]_{3}}^{\quad [2]_{6}\quad
LM}, \label{Ften}
\end{align}
\begin{align}
&F_{[1]_{3}[1]_{3}^{\ast} \quad[2,1,0]_{3}}^{\quad [2]_{6}\quad LM},
\quad F_{[1]_{3}[1]_{3}^{\ast} \quad[0]_{3}}^{\quad [2]_{6}\quad
LM}, \label{Flad}
\end{align}
which together with their conjugate (lowering) operators change the
number of bosons $N$ by two. The operators (\ref{Flad}) and their
conjugate counterparts are the ladder generators of $U(3,3)$
algebra.

In terms of Elliott's notations \cite{Elliott}
$\left(\lambda,\mu\right)$, we have $[2]_{3}=(2,0)$,
$[2]^{\ast}_{3}=[-2]_{3}=(0,2)$, $[210]_{3}=(1,1)$ and
$[0]_{3}=(0,0)$. The corresponding values of $L$ from the
$SU(3)\supset O(3)$ reduction rules are $L=0,2$ in both the $(2,0)$
and $(0,2)$ irreps, $L=1,2$ in the $(1,1)$ irrep and $L=0$ in the
$(0,0)$.

The number preserving operators transform according to the direct
product $[\chi]_{6}$ of the corresponding $U(6)$ representations
$[1]_{6}$ and $[1]_{6}^{\ast }$, namely
\begin{equation}
\lbrack 1]_{6}\times \lbrack 1]_{6}^{\ast}=[1,-1]_{6}+[0]_{6},
\label{prU6A}
\end{equation}
where $[1,-1]_{6}=[2,1,1,1,1,0]_{6}$ and $[0]_{6}=[1,1,1,1,1,1]_{6}$
is the scalar $U(6)$ representation. They generate the maximal
compact subgroup  $U(6)$ of $Sp(12,R)$.

The tensor operators
\begin{equation}
A_{[\lambda]_{3}[0]^{\ast}_{3}\quad [\lambda]_{3}}^{[1-1]_{6}\quad
LM} \simeq\frac{1}{\sqrt{2}} \sum_{m,k}C_{1m1k}^{LM}
p_{m}^{\dag}p_{k} \label{pten}
\end{equation}
\begin{equation}
A_{[0]_{3}[\lambda]^{\ast}_{3}\quad
[\lambda]^{\ast}_{3}}^{[1-1]_{6}\quad LM} \simeq\frac{1}{\sqrt{2}}
\sum_{m,k}C_{1m1k}^{LM}\ n_{m}^{\dag}n_{k} \label{nten}
\end{equation}
correspond to the generators of the $U_{p}(3)$ and
$\overline{U_{n}(3)}$ algebras, respectively. The operators with
$L=1$ represent the angular momentum components, whereas those with
$L=2$ correspond to the quadrupole momentum operators and together
they generate the one-fluid $SU_{\tau}(3)$ ($\tau=p,n$) algebra. The
tensors (\ref{pten}), (\ref{nten}) together with (\ref{Flad}) and
their conjugate counterparts, in turn, constitute the full set of
$U(3,3)$ generators.

The linear combination operators
\begin{equation}
A'^{LM}_{[\lambda]_{3}}=A^{[1-1]_{6}\quad
LM}_{[\lambda]_{3}[0]^{\ast}_{3}\quad
[\lambda]_{3}}-(-1)^{L}A^{[1-1]_{6}\quad
LM}_{[0]_{3}[\lambda]^{\ast}_{3}\quad [\lambda]_{3}} \label{u3z}
\end{equation}
generate the $U^{\ast}(3)$ algebra. The $SU^{\ast}(3)$ algebra is
obtained by excluding the operator $A'^{00}=N_{p}-N_{n} = M$ which
is the single generator of the $O(2)$ algebra, whereas the angular
momentum algebra $SO(3)$ is generated by the generators $A'^{1M}
\equiv L_{M}=L^{p}_{M}+L^{n}_{M}$ only. The operator $M$, counting
the difference between particle and holes, is also the first order
Casimir of $U(3,3)$ algebra and it decomposes the action space
$\mathcal{H}$ of the $Sp(12,R)$ generators to the ladder
${\mathcal{H}}_{\nu}$ subspaces of the boson representations of
$Sp(12,R)$ with $\nu=N_{p}-M_{n}=\pm 0, \pm 2, \pm 4,\ldots$
\cite{Sp2NRbr}.

Finally, the tensors
\begin{equation}
A^{[1-1]_{6} \quad LM}_{[1]_{3}[1]_{3}\quad
[\lambda]_{3}}\simeq\frac{1}{\sqrt{2}} \sum_{m,k}C_{1m1k}^{LM}\
p_{m}^{\dag}n_{k},
\end{equation}
\begin{equation}
A^{[1-1]_{6} \quad LM}_{[1]^{\ast}_{3}[1]^{\ast}_{3}\quad
[\lambda]_{3}}\simeq\frac{1}{\sqrt{2}} \sum_{m,k}C_{1m1k}^{LM}\
n_{m}^{\dag}p_{k}
\end{equation}
with $L=0,1,2$ and $M=-L,-L+1,...,L$ extend the $U_{p}(3)\otimes
\overline{U_{n}(3)}$ algebra to the $U(6)$ one.

In this way we have listed all the irreducible tensor operators in
respect to the reduction chain (\ref{NL}) that correspond to the
infinitesimal operators of the $Sp(12,R)$ algebra.

\section{Construction of the symplectic basis states of IVBM}

Next, we can introduce the tensor products
\begin{align}
&T^{([\chi_{1}]_{6}[\chi_{2}]_{6}) \quad
\omega[\chi]_{6}}_{[\lambda_{1}]_{3}[\lambda_{2}]_{3}
\quad\quad[\lambda]_{3} \quad \quad LM}= \notag \\
&\notag \\
&\sum T^{[\chi_{1}]_{6} \quad\quad L_{1}M_{1}
}_{[\lambda_{1}']_{3}[\lambda_{1}'']^{\ast}_{3} \ [\lambda_{1}]_{3}}
T^{[\chi_{2}]_{6} \quad\quad L_{2}M_{2}}_{[\lambda_{2}']_{3}[\lambda_{2}'']^{\ast}_{3} \ [\lambda_{2}]_{3}} \notag \\
&\notag \\
&\times C^{[\chi_{1}]_{6}\quad \quad \quad[\chi_{2}]_{6} \quad
\quad\quad
\omega[\chi]_{6}}_{[\lambda_{1}']_{3}[\lambda_{1}'']^{\ast}_{3}\quad
[\lambda_{2}']_{3}[\lambda_{2}'']^{\ast}_{3} \quad
[\lambda_{1}]_{3}[\lambda_{2}]_{3}}  \notag \\
&\notag \\
&\times C^{[\lambda]_{3}}_{[\lambda_{1}]_{3},[\lambda_{2}]_{3}}
C^{[\lambda_{1}]_{3} \quad [\lambda_{2}]_{3} \quad
[\lambda]_{3}}_{K_{1}L_{1} \ K_{2}L_{2} \quad KL} C^{L_{1} \quad
L_{2} \quad L}_{M_{1} \quad M_{2} \quad M} \label{tpr}
\end{align}
of two tensor operators $T^{[\chi_{i}]_{6} \quad\quad L_{i}M_{i}}
_{[\lambda'_{i}]_{3}[\lambda''_{i}]^{\ast}_{3} \
[\lambda_{i}]_{3}}$, which are as well tensors in respect to the
considered reduction chain. We use (\ref{tpr}) to obtain the
tensorial properties of the operators in the enveloping algebra of
$Sp(12,R),$ containing the products of the algebra generators. In
this particular case we are interested in the transition operators
between states differing by four bosons $T^{[4]_{6} \quad\quad
LM}_{[\lambda']_{3}[\lambda'']^{\ast}_{3} \ [\lambda]_{3}}$,
expressed in terms of the products of two operators $T^{[2]_{6}
\quad\quad L_{i}M_{i}}
_{[\lambda'_{i}]_{3}[\lambda''_{i}]^{\ast}_{3} \
[\lambda_{i}]_{3}}$. Making use of the decomposition (\ref{2NDec})
and the reduction rules in the chain (\ref{NL}), we list in Table 1
all the representations of the chain subgroups that define the
transformation properties of the resulting tensors.

\begin{table}[h]
\caption{Tensor products of two raising operators.} \label{T1}
\smallskip \centering
\begin{tabular}{|l|l|l|l|l|}
\hline
\begin{tabular}{l}
$\ \ \ \ [2]_{6}$ \\
$[ \lambda _{1}^{\prime }]_{3}[\lambda _{1}^{\prime \prime
}]_{3}^{\ast }$%
\end{tabular}
&
\begin{tabular}{l}
$\ \ \ \ [2]_{6}$ \\
$[ \lambda _{2}^{\prime }]_{3}[\lambda _{2}^{\prime \prime
}]_{3}^{\ast }$%
\end{tabular}
&
\begin{tabular}{l}
$\ \ \ \ [4]_{6}$ \\
$[ \lambda_{1}]_{3}[\lambda_{2}]_{3}^{\ast }$%
\end{tabular}
&
\begin{tabular}{l}
$U^{\ast }(3)$ \\
$\ [\lambda ]_{3}$%
\end{tabular}
&
\begin{tabular}{l}
$O(3)$ \\
$K;\ L$%
\end{tabular}
\\ \hline
$\ \ \ [2]_{3}[0]_{3}^{\ast}$ & $\ \ [2]_{3}[0]_{3}^{\ast }$ & $\ \
[4]_{3}[0]_{3}^{\ast }$ & $\ \ [4]_{3}$ & $0;0,2,4$ \\ \hline $\ \ \
[2]_{3}[0]_{3}^{\ast }$ & $\ \ [0]_{3}[2]_{3}^{\ast }$ & $\ \
[2]_{3}[2]_{3}^{\ast }$ & $\ [42]_{3}$ &
\begin{tabular}{l}
$2;2,3,4$ \\
$0;0,2$%
\end{tabular}
\\ \hline
$\ \ \ [2]_{3}[0]_{3}^{\ast }$ & $\ \ [0]_{3}[2]_{3}^{\ast }$ & $\ \
[2]_{3}[2]_{3}^{\ast }$ & $\ [321]_{3}$ & $1;1,2$ \\ \hline $\ \ \
[2]_{3}[0]_{3}^{\ast }$ & $\ \ [0]_{3}[2]_{3}^{\ast }$ & $\ \
[2]_{3}[2]_{3}^{\ast }$ & $\ \ [0]_{3}$ & $0;0$ \\ \hline $\ \ \
[0]_{3}[2]_{3}^{\ast }$ & $\ \ [0]_{3}[2]_{3}^{\ast }$ & $\ \
[0]_{3}[4]_{3}^{\ast }$ & $[-4]_{3}$ & $0;0,2,4$ \\ \hline
\end{tabular}
\end{table}

In order to clarify the role of the tensor operators introduced in
previous section as transition operators and to simplify the
calculation of their matrix elements, the basis for the Hilbert
space must be symmetry adapted to the  algebraic structure along the
considered subgroup chain (\ref{NL}). It is evident from
(\ref{Ften}) and (\ref{Flad}) that the basis states of the IVBM in
the $\mathcal{H}_{+}$ ($N-$even) subspace of the boson
representations of $Sp(12,R)$ can be obtained by a consecutive
application of the raising operators $F^{[\chi]_{6} \quad\quad
LM}_{[\lambda']_{3}[\lambda'']^{\ast}_{3} \ [\lambda]_{3}}$ on the
boson vacuum $\mid 0 \ \rangle$ (ground state) , annihilated by the
tensor operators $G^{[\chi]_{6} \quad\quad
LM}_{[\lambda']_{3}[\lambda'']^{\ast}_{3} \ [\lambda]_{3}}\mid 0 \
\rangle =0$ \ and $A^{[\chi]_{6} \quad\quad
LM}_{[\lambda']_{3}[\lambda'']^{\ast}_{3} \ [\lambda]_{3}}\mid 0 \
\rangle =0$.

Thus, in general a basis for the considered dynamical symmetry of
the IVBM can be constructed by applying the multiple symmetric
couplings (\ref{tpr}) of the raising tensors $T^{[2]_{6} \quad\quad
L_{i}M_{i}} _{[\lambda'_{i}]_{3}[\lambda''_{i}]^{\ast}_{3} \
[\lambda_{i}]_{3}}$ with itself - $[F\times \ldots \times \quad
F]^{[\chi]_{6} \quad\quad LM}_{[\lambda']_{3}[\lambda'']^{\ast}_{3}
\ [\lambda]_{3}}$. The possible $U^{\ast}(3)$ couplings are
enumerated by the set $[\lambda ]_{3}=\{[n_{1},n_{2},n_{3}]\equiv
(\lambda = n_{1}-n_{2},\mu = n_{2}-n_{3});n_{1}\geq n_{2} \geq
n_{3}\geq 0$ $\}$. We note that the integers $\{n_{i}\}$ can take
non-negative as well as negative values and hence correspond to
mixed irreps of $U^{\ast}(3)$ \cite{Flores}. The number of copies of
the operator $F$ in the symmetric product tensor $[N]_{6}$ is $N/2$,
where $N=N_{p}+N_{n}$. Each raising operator will increase the
number of bosons $N$ by two. Then, the resulting infinite basis can
be written as:
\begin{equation}
|[N]_{6};[N_{p}]_{3},[N_{n}]^{\ast}_{3};(\lambda,\mu); KLM \rangle,
\label{basis}
\end{equation}
where $[N]_{6}$, $[N_{p}]_{3}$ and $[N_{n}]^{\ast}_{3}$ denote the
irreducible representations of the $U(6)$, $U_{p}(3)$ and
$\overline{U_{n}(3)}$ groups respectively, while the quantum numbers
$KLM$ denote the basis of the irrep $(\lambda,\mu)$ of
$SU^{\ast}(3)$. By means of these labels, the basis states can be
classified in each of the two irreducible even ${\mathcal{H}}_{+}$
with $N=0,2,4,\ldots,$ and odd ${\mathcal{H}}_{-}$ with
$N=1,3,5,...,$  representations of $Sp(12,R)$.

\begin{widetext}

\begin{table}[h]
\caption{Symplectic classification of the $SU^{\ast}(3)$ basis
states.} \label{T2}
\smallskip \centering
\begin{tabular}{||l||llll|l|llll||}
\hline\hline $N\backslash \nu $ & $\cdots $ &
\multicolumn{1}{|l}{$6$} & \multicolumn{1}{|l}{$4$} &
\multicolumn{1}{|l|}{$2$} &  \  $\ \ 0$ & $-2$ &
\multicolumn{1}{|l}{$-4$} & \multicolumn{1}{|l}{$-6$} &
\multicolumn{1}{|l||}{$\cdots $} \\ \hline\hline $0$ &  &  &  &  &
$\ (0,0)$ &  &  &  &  \\ \cline{1-1}\cline{5-7} $2$ &  &  &
\begin{tabular}{l}
$F_{[2]_{3}[0]_{3}^{\ast }}^{[2]_{6}}$ \\
$\ \ \ \ \ \swarrow $%
\end{tabular}
& \multicolumn{1}{|l|}{%
\begin{tabular}{l}
$(2,0)$ \\
\ \
\end{tabular}%
} &
\begin{tabular}{l}
$(1,1)$ \\
$(0,0)$%
\end{tabular}
&
\begin{tabular}{l}
$(0,2)$ \\
\ \ \
\end{tabular}
& \multicolumn{1}{|l}{%
\begin{tabular}{l}
$F_{[0]_{3}[2]_{3}^{\ast }}^{[2]_{6}}$ \\
$\searrow $%
\end{tabular}%
} &  &  \\ \cline{1-1}\cline{4-8} $4$ &  & $F_{[1]_{3}[1]_{3}^{\ast
}}^{[2]_{6}}\downarrow $ &
\multicolumn{1}{|l}{%
\begin{tabular}{l}
$(4,0)$ \\
\\
\ \ \
\end{tabular}%
} & \multicolumn{1}{|l|}{%
\begin{tabular}{l}
$(3,1)$ \\
$(2,0)$ \\
\
\end{tabular}%
} &
\begin{tabular}{l}
$(2,2)$ \\
$(1,1)$ \\
$(0,0)$%
\end{tabular}
&
\begin{tabular}{l}
$(1,3)$ \\
$(0,2)$ \\
\
\end{tabular}
& \multicolumn{1}{|l}{%
\begin{tabular}{l}
$(0,4)$ \\
\\
\
\end{tabular}%
} & \multicolumn{1}{|l}{} &  \\ \cline{1-1}\cline{3-9} $6$ &
\begin{tabular}{l}
$A_{[1]_{3}^{\ast }[1]_{3}^{\ast }}^{[1-1]_{6}}$ \\
$\ \ \ \Longrightarrow $%
\end{tabular}
& \multicolumn{1}{|l}{%
\begin{tabular}{l}
$(6,0)$ \\
\\
\\
\
\end{tabular}%
} & \multicolumn{1}{|l}{%
\begin{tabular}{l}
$(5,1)$ \\
$(4,0)$ \\
\\
\
\end{tabular}%
} & \multicolumn{1}{|l|}{%
\begin{tabular}{l}
$(4,2)$ \\
$(3,1)$ \\
$(2,0)$ \\
\
\end{tabular}%
} &
\begin{tabular}{l}
$(3,3)$ \\
$(2,2)$ \\
$(1,1)$ \\
$(0,0)$%
\end{tabular}
&
\begin{tabular}{l}
$(2,4)$ \\
$(1,3)$ \\
$(0,2)$ \\
\
\end{tabular}
& \multicolumn{1}{|l}{%
\begin{tabular}{l}
$(1,5)$ \\
$(0,4)$ \\
\\
\
\end{tabular}%
} & \multicolumn{1}{|l}{%
\begin{tabular}{l}
$(0,6)$ \\
\\
\\
\
\end{tabular}%
} & \multicolumn{1}{|l||}{%
\begin{tabular}{l}
$A_{[1]_{3}[1]_{3}}^{[1-1]_{6}}$ \\
$\ \Longleftarrow $%
\end{tabular}%
} \\ \cline{1-1}\cline{3-9} $\vdots $ &  & \multicolumn{1}{|l}{ $\ \
\ \ \ \vdots $} & \multicolumn{1}{|l}{ $\ \ \ \ \vdots $} &
\multicolumn{1}{|l|}{ $\ \ \ \ \vdots $} &  $\ \ \ \ \vdots $ &  $\
\ \ \ \ \vdots $ & \multicolumn{1}{|l}{ $\ \ \ \ \vdots $} &
\multicolumn{1}{|l}{ $\ \ \ \ \vdots $} &
\multicolumn{1}{|l||}{}%
\end{tabular}
\end{table}
\end{widetext}

The $Sp(12,R)$ classification scheme for the $SU^{\ast}(3)$ boson
representations obtained by applying the reduction rules for the
irreps in the chain (\ref{NL}) for even value of the number of
bosons $N$ is shown on Table \ref{T2}. Each row (fixed $N$) of the
table corresponds to a given irreducible representation of the
$U(6)$ algebra, whereas the $SU^{\ast}(3)$ quantum numbers
$(\lambda,\mu)$ define the cells of the Table \ref{T2}. On the other
hand, the so called ladder representation of the non-compact algebra
$U(3,3)$ acts in the space of the boson representation of the
$Sp(12,R)$ algebra. Thus the ladder representations of $U(3,3)$
correspond to the columns (fixed value of $\nu$) of the Table
\ref{T2}. Note that along the columns the $SU^{\ast}(3)$ irreps
repeat each other except the ones corresponding to the first row for
each $N$.

Now, it is clear which of the tensor operators act as transition
operators between the basis states ordered in the classification
scheme presented on Table \ref{T2}. The operators
$F^{[2]_{6}\quad\quad LM}_{[1]_{3}[1]^{\ast}_{3}
\quad[\lambda]_{3}}$ give the transitions between two neighboring
cells $(\downarrow)$ from one column, while the $F^{[2]_{6}
\quad\quad LM} _{[2]_{3}[0]^{\ast}_{3} \quad[\lambda]_{3}}$
$(\swarrow )$ or $F^{[2]_{6} \quad\quad LM} _{[0]_{3}[2]^{\ast}_{3}
\quad[\lambda]_{3}}$ $(\searrow)$ ones change the column as well.
The tensors $A^{[1-1]_{6} \quad LM} _{[1]_{3}[1]_{3}
\quad[\lambda]_{3}}$ and $A^{[1-1]_{6} \quad LM}
_{[1]^{\ast}_{3}[1]^{\ast}_{3} \quad[\lambda]_{3}}$, acting within
the rows, change a given $SU^{\ast}(3)$ irrep to the neighboring one
on the left $(\Longleftarrow)$ and right $(\Longrightarrow)$,
respectively. The operators $A'^{LM}_{[210]_{3}}$ (\ref{u3z}), which
correspond to the $SU^{\ast}(3)$ generators do not change the
$SU(3)$ representations $(\lambda,\mu)$, but can change the angular
momentum $L$ inside it. The action of the tensor operators on the
$SU^{\ast}(3)$ vectors inside a given cell or between the cells of
Table \ref{T2}. is also schematically presented on it with
corresponding arrows, given above in parentheses.

\section{Matrix elements of the $U(3,3)$ ladder operators}

Physical applications are based on the correspondence of sequences
of $SU(3)$ vectors to sequences of collective states belonging to
different bands in the nuclear spectra. The above analysis permits
the  definition of the appropriate transition operators as
corresponding combinations of the tensor operators given in Sections
II and III.

In the present work we are interesting in the calculation of the
matrix elements of the $U(3,3)$ generators in appropriately chosen
symmetry-adapted basis. For this purpose we consider the following
reduction chain:
\begin{equation}
U(3,3) \supset U_{p}(3) \otimes \overline{U_{n}(3)} \supset
U^{\ast}(3) \supset SO(3),  \label{U33L}
\end{equation}
which is a part of (\ref{NL}). The basis is
\begin{equation}
|\nu;[N_{p}]_{3},[N_{n}]^{\ast}_{3};[\lambda]_{3}; KLM \rangle,
\label{u33bs}
\end{equation}
where $[\lambda]_{3}=(\lambda,\mu)$ and the new label $\nu$ denotes
the different $U(3,3)$ ladder representations. Note that the number
of bosons $N$ is not a good quantum number along the chain
(\ref{U33L}) and hence the $U(6)$ irrep label $[N]_{6}$ is
irrelevant and will be omitted in the further considerations.

The matrix elements of $U(3,3)$ generators can be calculated using
the fact that the Hilbert state space is the tensor product of the
$p-$ and $n-$boson representation spaces $[N_{p}]_{3}$ and
$[N_{n}]^{\ast}_{3}$, i.e.
\begin{equation}
|[N_{p}]_{3},[N_{n}]^{\ast}_{3};[\lambda]_{3} \rangle = |[N_{p}]_{3}
\rangle \otimes |[N_{n}]^{\ast}_{3} \rangle, \label{pntprs}
\end{equation}
coupled to good total $U^{\ast}(3)$ symmetry. Tensor operators in
the p-n space can be constructed by coupling tensors in the separate
spaces to good total $U^{\ast}(3)$ symmetry.

In the preceding sections we expressed all the symplectic generators
and the basis states as components of irreducible tensors in respect
to the reduction chain (\ref{NL}). Thus, for calculating of the
matrix elements of the $U(3,3)$ generators (which are a subset of
the symplectic generators), one can use the generalized
Wigner-Eckart theorem with respect to the $U_{p}(3)\otimes
\overline{U_{n}(3)}$ subgroup:
\begin{widetext}
\begin{align}
&\langle \nu;[N'_{p}]_{3},[N'_{n}]^{\ast}_{3};[\lambda']_{3};
K'L'M'|T^{\quad\quad\quad\quad lm}_{[\sigma']_{3}[\sigma'']_{3}\quad
[\sigma]_{3}}|\nu;[N_{p}]_{3},[N_{n}]^{\ast}_{3};[\lambda]_{3};
KLM \rangle \notag \\
\notag \\
&= \langle \nu;[N'_{p}]_{3},[N'_{n}]^{\ast}_{3};[\lambda']_{3};
K'L'||T^{\quad\quad\quad\quad lm}_{[\sigma']_{3}[\sigma'']_{3}\quad
[\sigma]_{3}}||\nu;[N_{p}]_{3},[N_{n}]^{\ast}_{3};[\lambda]_{3}; KL
\rangle C^{L'M'}_{LM,lm}. \label{ME}
\end{align}
\end{widetext}
Note that the $U(3,3)$ generators (\ref{Flad}) act within a given
ladder representation (fixed $\nu$) and change the number of bosons
$N$ by two, whereas the generators (\ref{Ften}) change the $U(3,3)$
irrep $\nu$ as well. The double-barred reduced matrix elements in
(\ref{ME}) are determined by the triple-barred matrix elements:
\begin{widetext}
\begin{align}
&\langle \nu;[N'_{p}]_{3},[N'_{n}]^{\ast}_{3};[\lambda']_{3};
K'L'||T^{\quad\quad\quad\quad lm}_{[\sigma']_{3}[\sigma'']_{3}\quad
[\sigma]_{3}}||\nu;[N_{p}]_{3},[N_{n}]^{\ast}_{3};[\lambda]_{3}; KL
\rangle \notag \\
\notag \\
&=\langle
\nu;[N'_{p}]_{3},[N'_{n}]^{\ast}_{3};[\lambda']_{3}|||T^{\quad\quad\quad\quad
lm}_{[\sigma']_{3}[\sigma'']_{3}\quad
[\sigma]_{3}}|||\nu;[N_{p}]_{3},[N_{n}]^{\ast}_{3};[\lambda]_{3}\rangle
C^{[\lambda]_{3} \quad [\sigma]_{3}\quad[\lambda']_{3}}_{KL \quad kl
\quad K'L'} \label{DBME}
\end{align}
\end{widetext}
where $C^{[\lambda]_{3} \quad [\sigma]_{3}\quad[\lambda']_{3}}_{KL
\quad kl \quad K'L'}$ are the $U(3)$ isoscalar factors and the
triple-barred  matrix elements depend only on the $U_{p}(3)$,
$\overline{U_{n}(3)}$ and $U^{\ast}(3)$ quantum numbers. Obviously,
for the evaluation of the matrix elements (\ref{ME}) of the $U(3,3)$
operators in respect to the chain (\ref{NL}) the knowledge of the
$U(3)$ IF as well as the reduced triple-barred matrix elements is
required.

We consider the $SO(3)$ reduced matrix element of the $U(3,3)$
ladder operator $F^{\quad \quad \quad \quad
lm}_{[1]_{3}[1]_{3}^{\ast} \quad[2,1,0]_{3}}\sim [p^{\dag}_{[1]_{3}}
\times n^{\dag}_{[1]_{3}^{\ast}}]^{\quad lm}_{[2,1,0]_{3}}$:
\begin{widetext}
\begin{align}
&\langle \nu;[N'_{p}]_{3},[N'_{n}]^{\ast}_{3};[\lambda']_{3};
K'L'||F^{\quad \quad \quad \quad lm}_{[1]_{3}[1]_{3}^{\ast}
\quad[2,1,0]_{3}}||\nu;[N_{p}]_{3},[N_{n}]^{\ast}_{3};[\lambda]_{3};
KL \rangle \notag \\
\notag \\
&=\langle
\nu;[N'_{p}]_{3},[N'_{n}]^{\ast}_{3};[\lambda']_{3}|||F^{\quad \quad
\quad \quad lm}_{[1]_{3}[1]_{3}^{\ast}
\quad[2,1,0]_{3}}|||\nu;[N_{p}]_{3},[N_{n}]^{\ast}_{3};[\lambda]_{3}\rangle
C^{[\lambda]_{3} \ \ [2,1,0]_{3}\quad[\lambda']_{3}}_{KL \quad \ kl
\quad \quad K'L'}. \label{DMEofF}
\end{align}
\end{widetext}
Since the operator under consideration acts on the separate $p-$ and
$n-$spaces, the reduced triple-barred matrix element can be
expressed as a product of the separate reduced triple-barred matrix
elements \cite{tme}:
\begin{widetext}
\begin{align}
&\langle
\nu;[N'_{p}]_{3},[N'_{n}]^{\ast}_{3};[\lambda']_{3}|||F^{\quad \quad
\quad \quad lm}_{[1]_{3}[1]_{3}^{\ast}
\quad[2,1,0]_{3}}|||\nu;[N_{p}]_{3},[N_{n}]^{\ast}_{3};[\lambda]_{3}\rangle \notag \\
\notag \\
&=\sum_{\rho_{p}\rho_{n}} \left\{
\begin{tabular}{llll}
$(N_{p},0)$ & $(1,0)$ & $(N_{p}^{\prime },0)$ & $\rho _{p}$ \\
$(0,N_{n})$ & $(0,1)$ & $(0,N_{n}^{\prime })$ & $\rho _{n}$ \\
$(N_{p},N_{n})$ & $(1,1)$ & $(N_{p}^{\prime },N_{n}^{\prime })$ & $1$ \\
$\ \ \ \ 1$ &  $\ \ 1$ &  $\ \ \ \ \ 1 $ &
\end{tabular}%
\right\} \langle [N'_{p}]_{3}|||p^{\dag}|||
[N_{p}]_{3}\rangle_{\rho_{p}} \langle
[N'_{n}]^{\ast}_{3}|||n^{\dag}|||[N_{n}]^{\ast}_{3}
\rangle_{\rho_{n}}, \label{TMEofF}
\end{align}
\end{widetext}
where $\{...\}$ stands for the $SU(3)$ $9-(\lambda,\mu)$ symbol. In
our case $\rho_{p}$ and $\rho_{n}$ are equal to 1, so there is no
sum in Eq.(\ref{TMEofF}). Taking into account that for the maximal
couplings (i.e. $N'_{p}=N_{p}+1$ and $N'_{n}=N_{n}+1$) the
corresponding $SU(3)$ $9-(\lambda,\mu)$ symbol is equal to 1, we
obtain for the reduced triple-barred matrix element
\begin{widetext}
\begin{align}
&\langle
\nu;[N_{p}+1]_{3},[N_{n}+1]^{\ast}_{3};[\lambda']_{3}|||F^{\quad
\quad \quad \quad lm}_{[1]_{3}[1]_{3}^{\ast}
\quad[2,1,0]_{3}}|||\nu;[N_{p}]_{3},[N_{n}]^{\ast}_{3};[\lambda]_{3}\rangle \notag \\
\notag \\
&=\sqrt{(N_{p}+1)(N_{n}+1)}, \label{TMEofF2}
\end{align}
\end{widetext}
where we have used the fact that in the case of vector bosons which
span the fundamental irrep $[1]$ of $u(n)$ algebra, the
$u(n)$-reduced matrix element of raising generators has the well
known form \cite{LeRo}.

The  $SO(3)$ reduced matrix element of the complimentary ladder
operator $G^{\quad \quad \quad \quad lm}_{[1]^{\ast}_{3}[1]_{3}
\quad[2,1,0]_{3}}\sim [p_{[1]^{\ast}_{3}} \times n_{[1]_{3}}]^{\quad
lm}_{[2,1,0]_{3}}$ of $U(3,3)$ algebra can be obtained from
Eq.(\ref{TMEofF}) and Eq.(\ref{TMEofF2}) simply by conjugation:
\begin{widetext}
\begin{align}
&\langle
\nu;[N_{p}-1]_{3},[N_{n}-1]^{\ast}_{3};[\lambda']_{3};K'L'||G^{\quad
\quad \quad \quad lm}_{[1]^{\ast}_{3}[1]_{3}
\quad[2,1,0]_{3}}||\nu;[N_{p}]_{3},[N_{n}]^{\ast}_{3};[\lambda]_{3};KL\rangle \notag \\
\notag \\
&\left(\langle
\nu;[N_{p}]_{3},[N_{n}]^{\ast}_{3};[\lambda']_{3}||F^{\quad \quad
\quad \quad lm}_{[1]_{3}[1]_{3}^{\ast}
\quad[2,1,0]_{3}}||\nu;[N_{p}-1]_{3},[N_{n}-1]^{\ast}_{3};[\lambda]_{3}\rangle\right)^{\ast} \notag \\
\notag \\
&=\sqrt{N_{p}N_{n}}C^{[\lambda]_{3} \quad [2,1,0]_{3}
\quad[\lambda']_{3}}_{KL \quad\quad kl \quad\quad K'L'}.
\label{DMEofG}
\end{align}
\end{widetext}
We want to point out that the isoscalar factors appearing in Eqs.
(\ref{DMEofF}) and (\ref{DMEofG}) are not known in general. A
computer code is available for their numerical evaluation
\cite{code}.

\section{B(E2) transition probabilities for the ground state band}

The most important point of the symplectic IVBM in the practical
applications to real nuclei is the identification of the
experimentally observed collective states of different bands with a
subset of the basis states from the symplectic extension given in
Table \ref{T2}. In general, an appropriate subset of $SU(3)$ states
are the so called "stretched" states \cite{stretched}. Their
domination is determined by the important role of the
quadrupole-quadrupole interactions in the collective excitations.
Thus, the most important $SU(3)$ states will be those with maximal
weight, i.e. those which have maximal eigenvalues of the second
order $SU(3)$ Casimir operator.

In the present approach we give as an example the evaluation of the
$B(E2)$ transition probabilities between the states of the ground
state band (GSB). For this purpose, we consider the following type
of stretched states $(\lambda,\mu) = (\lambda_{0}+k,\mu_{0}+k)$,
where $\lambda_{0}$ and $\mu_{0}$ fix the starting $SU^{\ast}(3)$
state built by $N_{0} = \lambda_{0}+ \mu_{0}$ bosons and $k$ is
changing. In our application, the integer number $k$ is related to
the angular momentum $L$ and gives rise to the collective bands.
Note that the presented type of the $SU^{\ast}(3)$ stretched states
are the states from the ladder representations (the columns of Table
\ref{T2}) of the $U(3,3)$ algebra. Hence an arbitrary transition
between these ladder states can be performed by the action of the
ladder operators of $U(3,3)$ or the tensor product operators from
the enveloping algebra of $Sp(12,R)$. For the GSB we chose
$N_{0}=0$, i.e. the initial $SU^{\ast}(3)$ state corresponding to
the ground state is $(\lambda_{0},\mu_{0})=(0,0)$. In this way, the
states of the GSB are identified with the $SU^{\ast}(3)$ multiplets
$(L,L)$. In order to visualize the correspondence under
consideration, we illustrate the selected subset of basis states in
Table \ref{sub}.

\begin{table}[h]
\caption{The subset of basis states (\protect\ref{u33bs}) associated
with the states of the GSB.}
\smallskip \centering
\begin{tabular}{||l||l|l|l|l|l|l||}
\hline\hline
$(\lambda ,\mu )$ & $(0,0)$ & $(2,2)$ & $(4,4)$ & $(6,6)$ & $(8,8)$ & $%
\ldots $ \\ \hline $\ \ L$ &  $\ \ 0$ &  $\ \ 2$ &  $\ \ 4$ &  $\ \
6$ &  $\ \ 8$ & $\ldots $
\\ \hline\hline
\end{tabular}
\label{sub}
\end{table}
As it was mentioned earlier, the vector bosons are considered as
elementary excitations or phonons that build different collective
states. Because of the latter, the same $U(3,3)$ irrep (i.e. the
same $SU^{\ast}(3)$ content in the pn-space as described above) is
associated with the states of the GSB for all nuclei under
consideration.

Transition probabilities are by definition $SO(3)$ reduced matrix
elements of transition operators $T^{E2}$ between the $|i\rangle
-$initial and $|f\rangle -$final collective states
\begin{equation}
B(E2;L_{i}\rightarrow L_{f})=\frac{1}{2L_{i}+1}\mid \langle \quad
f\parallel T^{E2}\parallel i\quad \rangle \mid ^{2}. \label{deftrpr}
\end{equation}
Using the tensorial properties of the $Sp(12,R)$ generators and the
mapping considered above, it is easy to define the $E2$ transition
operator between the states of the GSB band as:
%\begin{widetext}
\begin{align}
&T^{E2}=e[A'^{20}_{[210]_{3}} +\notag\\
\notag\\
&\theta ([F\times F]^{\quad \quad \quad \quad
20}_{[2]_{3}[2]^{\ast}_{3}\quad [420]_{3}}+[G\times G]^{\quad \quad
\quad \quad 20}_{[2]^{\ast}_{3}[2]_{3}\quad [420]_{3}})],
\label{te2}
\end{align}
%\end{widetext}
where the first tensor operator is the $SU^{\ast}(3)$ quadrupole
operator and actually changes only the angular momentum with $\Delta
L=2$ within a given irrep $(\lambda,\mu)$.

The tensor product
%\begin{widetext}
\begin{align}
&[F\times F]^{\quad \quad \quad \quad
20}_{[2]_{3}[2]^{\ast}_{3}\quad [420]_{3}} \notag \\
\notag \\
&=\sum C^{[420]_{3}}_{[2]_{3},[2]^{\ast}_{3}} C^{(2,0) (0,2)
(2,2)}_{\quad 2 \quad 2 \quad 2} C^{20}_{20,20} \notag \\
\notag \\
&\times F^{\quad \ \quad \quad 20}_{[2]_{3}[0]^{\ast}_{3}\quad
[2]_{3}} F^{\quad \ \quad \quad 20}_{[0]_{3}[2]^{\ast}_{3}\quad
[-2]_{3}} \label{FF}
\end{align}
%\end{widetext}
of the rasing generators of $Sp(12,R)$ changes the number of bosons
by $\Delta N=4$ and $\Delta L=2$.

For the $SO(3)$ reduced matrix element of the tensor product
$[F\times F]^{\quad \quad \quad \quad
20}_{[2]_{3}[2]^{\ast}_{3}\quad [420]_{3}}$ between the states of
the GSB we obtain
\begin{widetext}
\begin{align}
&\langle
0;[N_{p}+2]_{3},[N_{n}+2]^{\ast}_{3};[\lambda']_{3};K'=0L'||[F\times
F]^{\quad \quad \quad \quad
20}_{[2]_{3}[2]^{\ast}_{3}\quad [420]_{3}}||0;[N_{p}]_{3},[N_{n}]^{\ast}_{3};[\lambda]_{3};K=0L\rangle \notag \\
\notag \\
&=C^{[\lambda]_{3} \ [4,2,0]_{3} \ [\lambda']_{3}}_{KL \quad kl
\quad \ K'L'}\sum_{\rho_{p}\rho_{n}} \left\{
\begin{tabular}{llll}
$(N_{p},0)$ & $(2,0)$ & $(N_{p}+2,0)$ & $\rho _{p}$ \\
$(0,N_{n})$ & $(0,2)$ & $(0,N_{n}+2)$ & $\rho _{n}$ \\
$(N_{p},N_{n})$ & $(2,2)$ & $(N_{p}+2,N_{n}+2)$ & $1$ \\
$\ \ \ \ 1$ &  $\ \ 1$ &  $\ \ \ \ \ 1 $ &
\end{tabular}%
\right\} \langle [N_{p}+2]_{3}|||F||| [N_{p}]_{3}\rangle_{\rho_{p}}
\langle [N_{n}+2]^{\ast}_{3}|||F|||[N_{n}]^{\ast}_{3}
\rangle_{\rho_{n}} \notag \\
\notag \\
&=\sqrt{(N_{p}+1)(N_{p}+2)(N_{n}+1)(N_{n}+2)}C^{[\lambda]_{3} \
[4,2,0]_{3} \ [\lambda']_{3}}_{KL \quad kl \quad \ K'L'},
\label{RMEFF}
\end{align}
\end{widetext}
where again for the case of the maximal couplings
$\rho_{p}=\rho_{n}=1$ and hence there is no sum in Eq.(\ref{RMEFF})
and the $SU(3)$ 9-$(\lambda,\mu)$ coefficient is equal to 1. In
Eq.(\ref{RMEFF}) we have used the standard recoupling technique for
two coupled $U(3)$ tensors \cite{Recoupling}:
\begin{align}
&\langle [N_{p}+2]_{3}|||F|||[N_{p}]_{3} \rangle \notag\\
\notag\\
&=U([N_{p}]_{3};[1]_{3};[N_{p}+2]_{3};[1]_{3}|[N_{p}+1]_{3};[2]_{3}) \notag\\
\notag\\
&\times \langle [N_{p}+2]_{3}|||p^{\dag}_{[1]_{3}}|||[N_{p}+1]_{3}
\rangle \notag\\
\notag\\
&\times \langle [N_{p}+1]_{3}|||p^{\dag}_{[1]_{3}}|||[N_{p}]_{3}
\rangle,
\end{align}
where $U(\ldots)$ denotes the $U(3)$ Racah coefficient, which for
maximal couplings is equal to 1.

Similarly, for the $SO(3)$ reduced matrix element of the tensor
product $[G\times G]^{\quad \quad \quad \quad
20}_{[2]^{\ast}_{3}[2]_{3}\quad [420]_{3}}$ we obtain
\begin{widetext}
\begin{align}
&\langle
0;[N_{p}-2]_{3},[N_{n}-2]^{\ast}_{3};[\lambda']_{3};K'=0L'||[G\times
G]^{\quad \quad \quad \quad
20}_{[2]^{\ast}_{3}[2]_{3}\quad [420]_{3}}||0;[N_{p}]_{3},[N_{n}]^{\ast}_{3};[\lambda]_{3};K=0L\rangle \notag \\
\notag \\
&=\sqrt{N_{p}(N_{p}-1)N_{n}(N_{n}-1)}C^{[\lambda]_{3} \ [4,2,0]_{3}
\ [\lambda']_{3}}_{KL \quad kl \quad \ K'L'}. \label{RMEGG}
\end{align}
\end{widetext}

Finally, we calculate the matrix element of the quadrupole operator
$A'^{20}_{[210]_{3}}$ using the fact that it is an $SU^{\ast}(3)$
generator. So, the  Wigner-Eckart theorem is applied in respect to
the $SU^{\ast}(3)$ subgroup
\begin{widetext}
\begin{align}
&\langle 0;[N'_{p}]_{3},[N'_{n}]^{\ast}_{3};(N'_{p},N'_{n});0L-2||
A'^{20}_{[210]_{3}}||0;[N_{p}]_{3},[N_{n}]^{\ast}_{3};(N_{p},N_{n});0L\rangle \notag \\
\notag \\
&=\delta _{N_{p}N'_{p}}\delta _{N_{n}N'_{n}}\sum\limits_{\rho
=1,2}C_{\quad L-2 \ \ \quad 2 \quad \quad L}^{(N'_{p},N'_{n})\
(1,1)\ \rho (N_{p},N_{n})} \langle
(N'_{p},N'_{n})|||A'^{20}_{[210]_{3}}|||(N_{p},N_{n})\rangle_{\rho}.
\label{RMEA}
\end{align}
\end{widetext}
The reduced triple-barred matrix elements are well known and are
given for $\rho=1$ by \cite{RRAnn}
\begin{align}
&\langle
(\lambda=N_{p},\mu=N_{n})|||A'^{20}_{[210]_{3}}|||(\lambda=N_{p},\mu=N_{n})\rangle_{1}
\notag \\
&= \left\{
\begin{array}{c}
g_{\lambda \mu }, \ \ \mu =0 \\
-g_{\lambda \mu }, \  \mu \neq 0%
\end{array}
\right. \label{RTMEofA}
\end{align}
where
\begin{equation}
g_{\lambda \mu }=2\left(\frac{\lambda ^{2}+\mu ^{2}+\lambda \mu +3\lambda +3\mu }{%
3}\right)^{1/2}
\end{equation}
and the phase convention is chosen to agree with that of Draayer and
Akiyama \cite{code}. For $\rho=2$ we have $\langle
(\lambda,\mu)|||A'^{20}_{[210]_{3}}|||(\lambda,\mu)\rangle_{2}=0$.

With the help of the above analytic expressions (\ref{RMEFF}),
(\ref{RMEGG}) and (\ref{RMEA}) for the matrix elements of the tensor
operators forming the $E2$ transition operator we can calculate the
transition probabilities (\ref{deftrpr}) between the states of the
ground state band as attributed to the $SU^{\ast}(3)$
symmetry-adapted basis states of the model (\ref{u33bs}). All the
required $U(3)$ IF's are numerically obtained using the computer
code \cite{code}.

\section{Application to real nuclei}

In order to test the model predictions following from our
theoretical considerations we apply the theory to real nuclei
exhibiting axially asymmetric features for which there is enough
available experimental data for the transition probabilities between
the states of the ground bands from the $A \sim 100$ and $A \sim
190$ mass regions. The application actually consists of fitting the
two parameters $e$ and $\theta$ of the transition operator $T^{E2}$
(\ref{te2}) to experiment for each isotope.

\begin{figure}[h]\centering
\includegraphics[width=80mm]{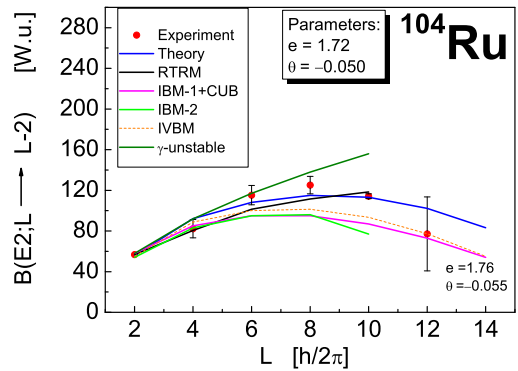}
\caption{(Color online) Comparison of theoretical and experimental
values for the $B(E2)$ transition probabilities in $^{104}Ru$. The
theoretical results of IBM-1 with a cubic term included, IBM-2,
Rigid Triaxial Rotor Model, and $\gamma$-unstable model of Wilets
and Jean are also shown.} \label{Ru104}
\end{figure}

As a first example we consider the intraband $B(E2)$ transitions in
the GSB for the nucleus $^{104}Ru$, which was assumed to possess
transitional properties between the $\gamma$-soft ($O(6)$ limit) and
$\gamma$-rigid ($SU^{\ast}(3)$ limit) structures \cite{PSDiep},
\cite{Diep}. The $^{96-108}Ru$ isotopes have also been described
within the framework of IBM-1 as transitional between $U(5)$ and
$O(6)$ limits \cite{IBMRu}, whereas in the Generalized Collective
Model these nuclei are described as transitional between spherical
and triaxial with a prolate onset for $^{96}Ru$ \cite{GCMRu}. The
experimental data \cite{IBMts} for the $B(E2)$ transition
probabilities between the states of the GSB are compared with the
corresponding theoretical results of the symplectic IVBM in Figure
\ref{Ru104}. For comparison, the theoretical predictions of the
IBM-1 \cite{IBMts}, including a cubic term producing a stable
triaxial minimum,  those of the IBM-2 \cite{IBM2Ru104}, Rigid
Triaxial Rotator Model (RTRM) \cite{Toki}, and $\gamma$-unstable
model of Wilets and Jean \cite{WJ} are also shown. From the figure
one can see that all models presented reproduce the general trend of
the experimental data, but nevertheless the latter lie between the
predictions of the $\gamma$-unstable and $\gamma$-rigid models,
suggesting a more complex and intermediate situation between these
two structures. Note the identical curves for IBM-1 and IBM-2 up to
$L \simeq 8$. With a slightly modified values of the parameters
$\theta $ and $e$, the IVBM results become very similar to those of
IBM, which is also illustrated in the Figure \ref{Ru104} (dashed
curve).

Next, we present the theoretical results for some nuclei from the $A
\sim 190$ mass region. The Pt-Os region is traditionally considered
within the IBM-1 framework to be a good example for the transition
between $SU(3)$ and $O(6)$ \cite{IBM-OsPt}. A number of theoretical
calculations \cite{Bonche}, \cite{CCQH}, \cite{Robledo},
\cite{Nomura1}, \cite{Nomura2} predict a tiny region of triaxiality
between the prolate and oblate shapes in this mass region. Recent
self-consistent Hartree-Fock-Bogoliubov calculations \cite{Robledo}
with Gogny D1S and Skyrme SLy4 forces predict that the prolate to
oblate transition takes place at neutron number $N=116$ ($^{192}$Os,
$^{194}$Pt).

\begin{figure}[h]\centering
\includegraphics[width=80mm]{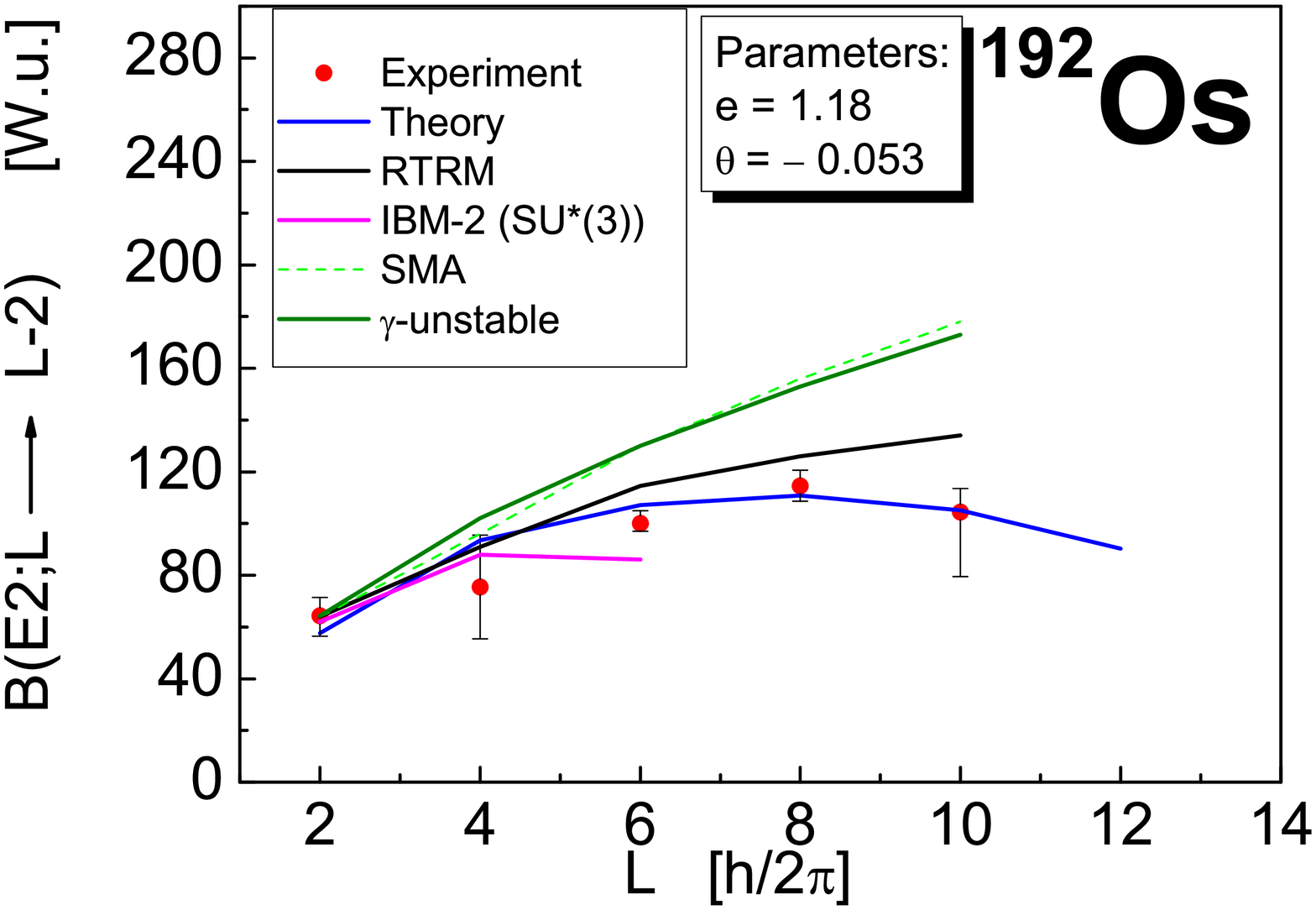}
\caption{(Color online) Comparison of theoretical and experimental
values for the $B(E2)$ transition probabilities in $^{192}Os$. The
theoretical results of the Rigid Triaxial Rotor Model, IBM-2  in its
$SU^{\ast}(3)$ limit, sextic and Mathieu approach (SMA), and
$\gamma$-unstable model are also shown.} \label{Os192}
\end{figure}

In Figure \ref{Os192}, the experimental $B(E2)$ values for
transitions between the members of the GSB in $^{192}Os$ are
compared with the theoretical results of IVBM, IBM-2 \cite{Walet}
($SU^{\ast}(3)$ limit), RTRM \cite{Toki}, sextic and Mathieu
approach (SMA) \cite{Raduta}, and $\gamma$-unstable model of Wilets
and Jean \cite{WJ}. One can see a slight reduction of the
collectivity with the increasing spin well described by the IVBM,
whereas the RTRM, SMA, and $\gamma$-unstable model of Wilets and
Jean overestimate the observed experimental data.

\begin{figure}[h]\centering
\includegraphics[width=80mm]{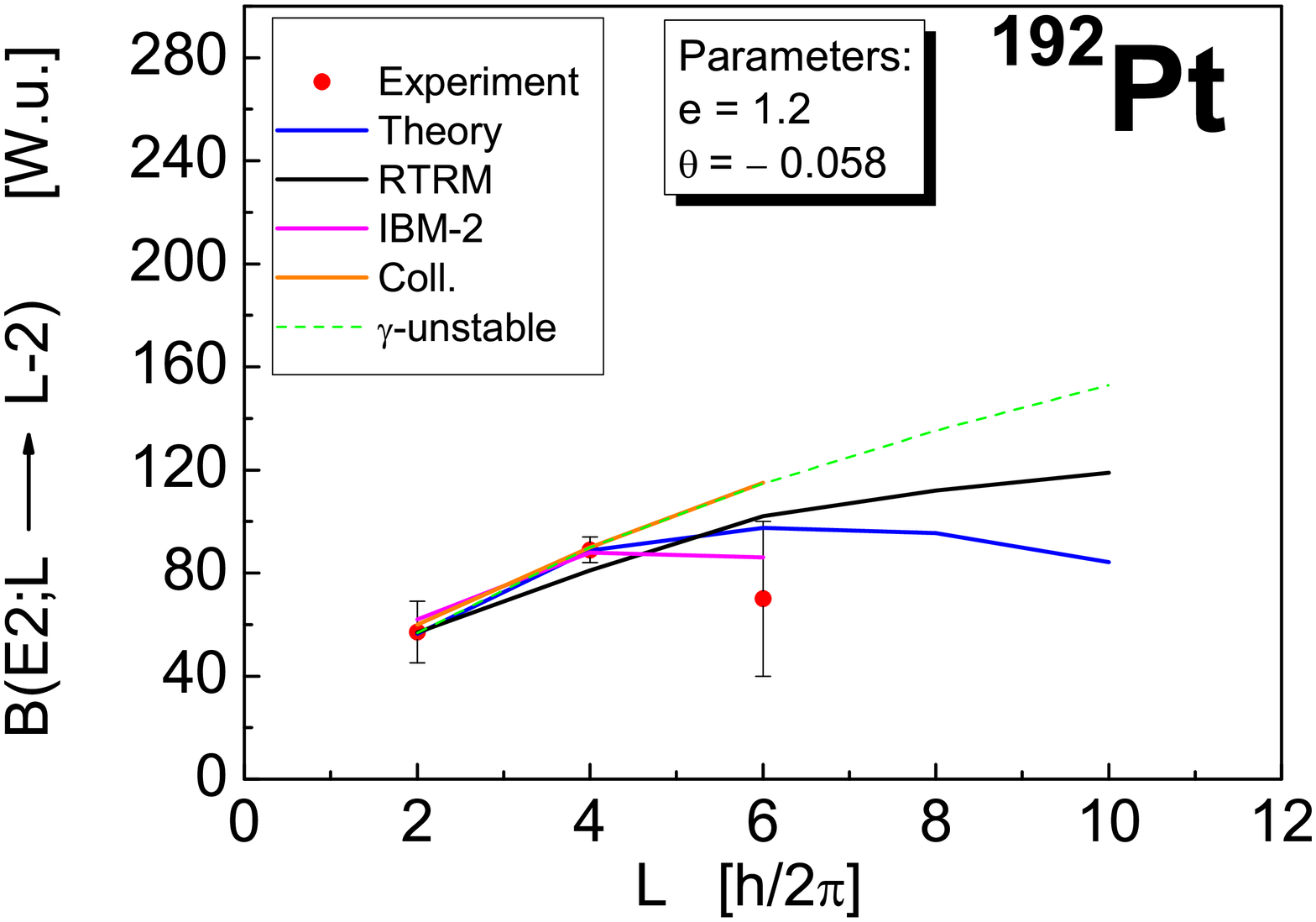}
\caption{(Color online) Comparison of theoretical and experimental
values for the $B(E2)$ transition probabilities in $^{192}Pt$. The
theoretical results of the Rigid Triaxial Rotor Model, IBM-2,
Quadrupole Collective Model, and $\gamma$-unstable model are also
shown.} \label{Pt192}
\end{figure}

\begin{figure}[h]\centering
\includegraphics[width=80mm]{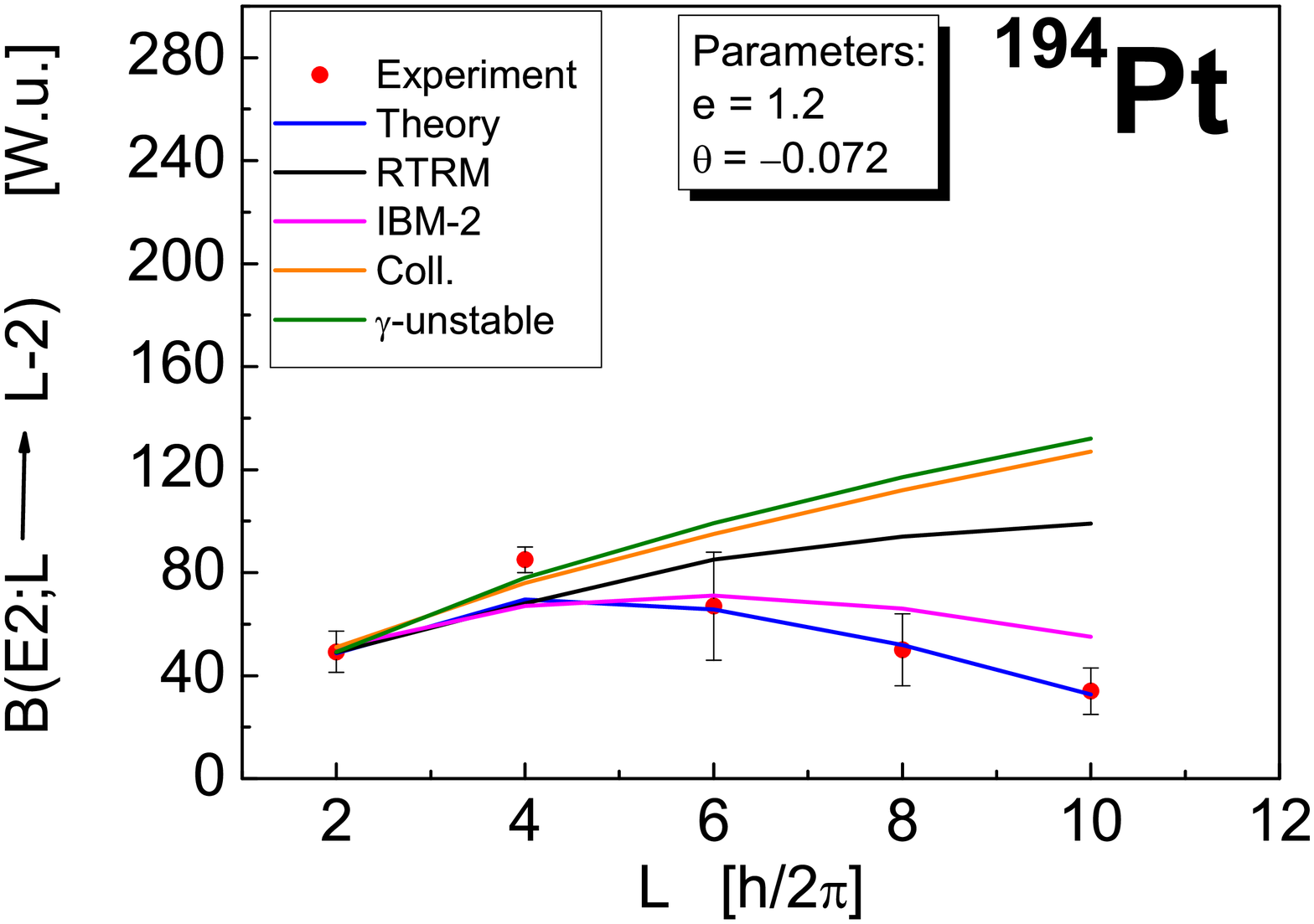}
\caption{(Color online) Comparison of theoretical and experimental
values for the $B(E2)$ transition probabilities in $^{194}Pt$. The
theoretical results of the Rigid Triaxial Rotor Model, IBM-2,
Quadrupole Collective Model, and $\gamma$-unstable model are also
shown.} \label{Pt194}
\end{figure}

Next, the experimental $B(E2)$ values \cite{Nomura2} between the
states of the GSB in $^{192}Pt$ and $^{194}Pt$ isotopes are shown in
Figures \ref{Pt192} and \ref{Pt194}, respectively, compared with the
theoretical predictions of IVBM from one side, and those of IBM-2
\cite{Nomura2}, RTRM \cite{Toki}, the Quadrupole Collective Model
(Coll.) \cite{Nomura2}, and $\gamma$-unstable model \cite{WJ} from
another. The reduction in the $B(E2)$ values with increasing spin is
well described by the IVBM in the two nuclei, compared to the
predictions of other collective models.

From Figs. \ref{Ru104}$-$\ref{Pt194} one can see that the IVBM
describes the $B(E2)$ transitions probabilities between the
collective states of the GSB in the four considered even-even nuclei
rather well. At this point we want to make some comments concerning
the two parameters $e$ and $\theta$. Detailed analysis shows that
the two main types of $B(E2)$ behavior - the enhancement or the
reduction of the $B(E2)$ values - can be described within the
present approach. The change of the values of the parameter $e$
affects mainly the scale. The coefficient in front of the second
term in Eq.(\ref{te2}) is about of two orders of magnitude smaller
than the $SU(3)$ contribution to the transition operator
(\ref{te2}), but its role in reproducing the correct behavior of the
transition probabilities between the states of the GSB is very
important. At $\theta = 0$ the theory gives a very specific, almost
"linear", behavior of the $B(E2)$ values. For $\theta < 0 $, with
the increasing of the absolute value of the parameter $\theta$ - the
theoretical curve goes from that of enhanced $B(E2)$ values (which
is an indication for the enhanced collectivity in the high angular
momentum domain) to the case of the well-known "cutoff effect",
which is a characteristic feature of all $SU(3)$-based calculations.

\begin{figure}[h]\centering
\includegraphics[width=80mm]{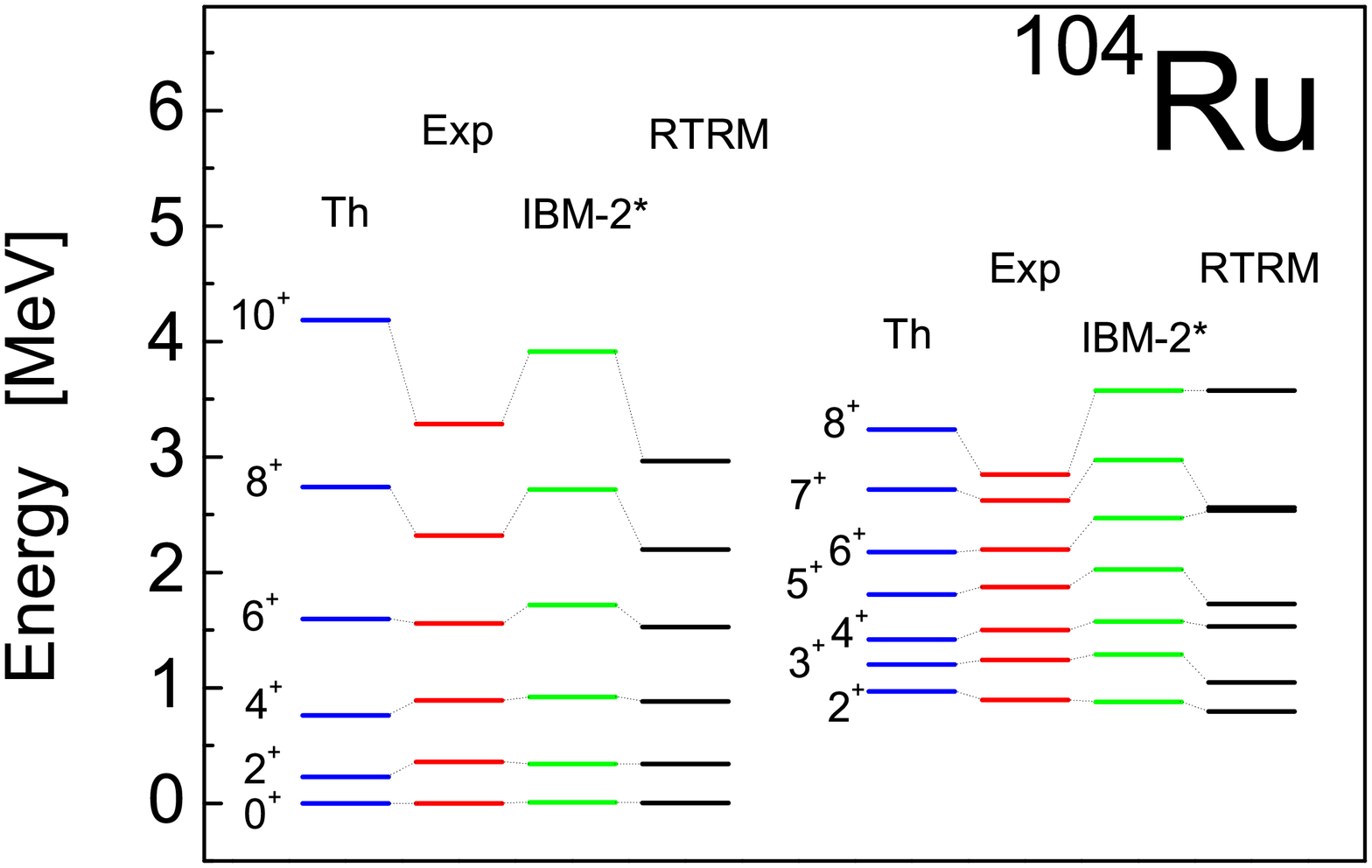}
\caption{(Color online) Excitation energies of the GSB and $\gamma$
band in $^{104}Ru$, compared with the experimental data and the
predictions of IBM-2 in its $SU^{\ast}(3)$ limit and RTRM. The
values of the model parameters are $a_{1} = 0.2155$ MeV, $b
=-0.0098$ MeV, $a_{3} =-0.0002$ MeV, and $b_{3} =0.0387$ MeV.}
\label{Ru104e}
\end{figure}

Being a group of dynamical symmetry, the $Sp(12,R)$ through its
reduction given by Eq.(\ref{NL}) determines the type of spectra
(obtained at fixed values of the model parameters in the
Hamiltonian) of different nuclei that it can describe. As an
illustration, in Fig. \ref{Ru104e} we show the theoretical results
for the excitation energies of the ground and $\gamma$ bands in
$^{104}Ru$, compared with the experimental data and the predictions
\cite{Ru104e} of IBM-2 in its $SU^{\ast}(3)$ limit and RTRM, both of
which incorporate $\gamma$-rigid structures. The states of the
$\gamma$ band are associated with the stretched states from the $\nu
= -2$ irrep of $U(3,3)$. (Detailed comparison of the energy spectra
obtained in the present approach for some even-even nuclei, assumed
to be axially asymmetric, with experiment will be given elsewhere.)
The Hamiltonian used in our calculation, expressed as a linear
combination of the Casimir operators along the chain (\ref{NL}), is
of the form
\begin{equation}
H=a_{1}M^{2} + b(N_{n}^{2}-N_{p}^{2}) + a_{3}C_{2}[SU^{\ast}(3)] +
b_{3}C_{2}[SO(3)]. \label{HU33}
\end{equation}
The values of the model parameters are determined by fitting the
energies of the ground and $\gamma$ bands in $^{104}Ru$ to the
experimental data \cite{exp}, using a $\chi^{2}$-procedure. From the
Fig. \ref{Ru104e} we see that the IVBM results are very similar to
the ones predicted by the IBM-2. The RTRM gives better description
of the collective states of the GSB, while for the $\gamma$ band it
gives pronounced  $\gamma$-rigid doublet structure not observed in
experiment. The latter shows more regular spacings of the states in
the $\gamma$ band, reasonably well reproduced by both the IVBM and
IBM-2.

The results obtained for both the $B(E2)$ transition probabilities
between the collective states of the GSB in the even-even nuclei
under consideration and the energy levels of the GSB and $\gamma$
band in $^{104}Ru$ prove the correct mapping of the basis states to
the experimentally observed ones. We recall the transitional
character of the nucleus $^{104}Ru$ between $\gamma$-unstable
($O(6)$ limit) and $\gamma$-rigid ($SU^{\ast}(3)$ limit) in terms of
the IBM. In this way the theoretical results obtained within the
framework of IVBM suggest the range of the applicability of the
present approach and reveal its relevance in the description of
nuclei that exhibit axially asymmetric features in their spectra.

\section{Conclusions}

In the present paper we investigated the tensor properties of the
algebra generators of $Sp(12,R)$ with respect to the reduction chain
(\ref{NL}). $Sp(12,R)$ is the group of dynamical symmetry of the
IVBM. The basis states of the model are also classified by the
quantum numbers corresponding to the irreducible representations of
the subgroups from the chain. The action of the symplectic
generators as transition operators between the basis states is
analyzed. The matrix elements of the $U(3,3)$ ladder operators in
the so obtained symmetry-adapted basis are given.

The $U(3,3)$ limit of the symplectic IVBM is further tested on the
more complicated and complex problem of reproducing the $B(E2)$
transition probabilities between the states of the ground band in
some even-even nuclei from the $A=190$ and $A=190$ mass regions
assumed by many authors to be axially asymmetric. In developing the
theory the advantages of the algebraic approach are used for the
assignment of the basis states to the experimentally observed states
of the collective bands and the construction of the $E2$ transition
operator as linear combination of tensor operators representing the
generators of the subgroups of the respective chain. This allows the
application of a specific version of the Wigner-Eckart theorem and
consecutively leads to analytic results for their (reduced) matrix
elements in the $U(3,3)$ symmetry-adapted basis that give the
transition probabilities.

In the application to real nuclei, the parameters of the transition
operator are evaluated in a fitting procedure for GSB of the
considered nuclei. The $B(E2)$ transition probabilities between
collective states of the ground state band in $^{104}Ru$,
$^{192}Os$, $^{192}Pt$, and $^{194}Pt$ isotopes are calculated and
compared with the experimental data and some other collective models
that accommodate the $\gamma$-rigid or $\gamma$-soft structures. The
experimental data for the presented examples are reproduced rather
well, although the results are very sensitive to the values of the
model parameters.

Being a group of dynamical symmetry, the $Sp(12,R)$ through its
reduction given by Eq.(\ref{NL}) determines the type of spectra
(obtained at fixed values of the model parameters in the
Hamiltonian) of different nuclei that it can describe. The excited
states of the GSB and $\gamma$ band in the transitional nucleus
$^{104}Ru$ are calculated within the IVBM using a four parameter
Hamiltonian, expressed as a linear combinations of the Casimir
operators along the dynamical chain (\ref{NL}) and compared with the
experimental data and the predictions of IBM-2 in its $SU^{\ast}(3)$
limit and RTRM, both of which incorporate $\gamma$-rigid structures.
The structure of the two bands is reasonably well described by the
present approach.

Summarizing, the results obtained for both the $B(E2)$ transitions
probabilities between the collective states of the GSB in the
even-even nuclei under consideration and the energy levels of the
GSB and $\gamma$ band in $^{104}Ru$ prove the correct mapping of the
basis states to the experimentally observed ones and reveal the role
of the symplectic symmetries in the description of nuclei,
exhibiting axially asymmetric features in their spectra.

\section*{Acknowledgment}

This work was supported by the Bulgarian National Foundation for
scientific research under Grant Number DID-$02/16/17.12.2009$.

\end{document}